       \documentstyle [12pt]   {article}
 \evensidemargin\oddsidemargin
\begin{document}
\count0 = 1
 \title{{ Quantum Measurements and Information \\
 Restrictions in Algebraic QM   \\ }}
\small\author{S.N.Mayburov \\ 
Lebedev Inst. of Physics\\
Leninsky Prospect 53, Moscow, Russia, 117924\\
E-mail :\quad   mayburov@sci.lpi.msk.su}
\date {}
\maketitle
\begin{abstract}

  Information-Theoretical restrictions on the systems measurements
and the information acquisition are  applied to Quantum Measurements Theory.
For the measurement of quantum object S  by  the information system $O$
such restrictions  are  described by the formalism of
 restricted states which obtained from 
 the agreement
with Shr$\ddot {o}$dinger dynamics and measurement statistics.
 The analogous  restrictions derived in Algebraic QM  formalism
 from the consideration of Segal algebra $\cal U_O$  of $O$ observables;
the resulting $O$  restricted states $\{\xi^O_i\}$ set
is defined as $ \cal U_O$  dual space. From
 Segal theorem for associative  (sub)algebras  it's shown
 that  $\xi^O_j$ describes the  random 'pointer'
outcomes $q_j$  observed by $O$ in the individual events.

\end{abstract}
\vspace{6mm}
$ \quad \quad $ {\bf Key Words: \, Quantum Measurements, Information,
  Observable algebras}  
\small { \quad 
\vspace{10mm}
%
%
%

\section  {Introduction}

There are still  several unsettled problems
 concerning the interpretation of 
 Quantum Mechanics (QM), and the majority of them
 involves to some extent the quantum measurement processes
(Jauch,1968; Aharonov,1981). 
 The oldest and most prominent of them is probably
  the State Collapse or Objectification
   Problem  (D'Espagnat,1990; Busch,1996).
In this paper we shall analyze the quantum  measurements  within the
 Information-Theoretical framework which  considers some  of their important
aspects.
Indeed,   the
  measurement of any kind is   the  reception of data about
 the observed system S parameters   by another  system $O$ (Observer) 
 (Guilini,1996; Duvenhage,2002).  
Therefore   the  possible 
  Information-Theoretical restrictions on the information 
 transfer from  S to $O$  can influence
on the observed measurement effects  (Svozil,1993).
%
In the model regarded here
 $O$ is   the information gaining and utilizing system 
(IGUS);  it processes and memorizes the  information acquired 
 as the result of
 S interactions   with the measuring system (MS) which element is $O$.
 We  assume  that QM description is applicable
both for  a microscopic    and  macroscopic
objects, 
this is the standard assumption of  Quantum Measurement theory, 
despite that it's not quite well founded (Busch,1996).
 In particular, in our approach  $O$ state  is described 
by the  quantum state  $\rho$ 
  relative to another  observer $O'$ (Rovelli,1995;Bene,2000).
In this approach $O$  can be  either a human brain
 or some automatic device,  in all cases it's
the system  
  which final  state correlates with the input data.

Even if 
  S measurement by $O$ is described by the evolution
of  MS state $\rho _{MS}$ 
relative to some external $O'$,  our aim is
 to calculate    
  S information  acquired  by $O$ itself.  In general  $\rho_{MS}$
ansatz isn't sufficient for that,
 and the description of MS state  relative to $O$  
 should be regarded in the  self-description framework
of Information Theory (Svozil,1993).
Earlier the  systems self-description
was investigated in the context of
 the general problem of mathematical  Self-reference 
 (Finkelstein,1988; Mittelstaedt,1998).
It was argued that the self-description of an arbitrary system 
is always incomplete; this result often interpreted as the
analog of G$\it\ddot{o}$del Theorem for Information Theory (Svozil,1993).
 Recently the self-description
 in the measurement process,  called also the  measurement from inside,
 was considered in the  formalism of inference maps $M_O$  
 connecting MS and $O$ states $\rho_{MS}, R_O$ (Breuer,1996). It
 leads to some general results for the quantum measurements, 
 but by itself doesn't permit to derive $M_O$  
  from the  first  QM principles.
In this paper we argue that the  inference map 
or arbitrary systems can be  
calculated in Algebraic QM framework which exploits
    Segal  algebras of observables - i.e. $C^*$- algebras formalism.
 (Emch,1972). In this formalism MS  state $\xi^{MS}$ is defined on MS 
observables algebra $\cal{U}$,
$O$ state $\xi^O$  
  is defined on $O$ observables subalgebra $\cal{U}_O$ which is 
the restricted set of MS observables.
It will be shown that for  MS in the pure state,  $\xi^O$ is  
the stochastic state which describes the random outcomes
 in the measurement of S states superpositions and this effect
can be interpreted as the subjective state collapse observed by $O$.  
Earlier the analogous inference map was obtained phenomenologically
in  the formalism of doublet states (Mayburov,2001,2004).
Alternatively  this 
 results can be interpreted also as the consequences of principal 
restrictions on
the information transfered from S to $O$. 



%
We should stress that  the observer's consciousness 
doesn't play any role in our theory and isn't referred to directly anywhere
(London,1939).
%
 The terms 'perceptions', 'impressions' used here
 are  defined  in strictly physical  terms.
In our model the perception is  
 the acquisition of some information by $O$, i.e. the change
of $O$  state;   a different
 $O$ impressions are associated with a different  $O$ states;
 their properties  are discussed  also in the final chapter.
 The  states  of objects defined in  $O$ reference frame (RF)
 (or other  $O'$ RF)
   are referred to also as 'S  state for $O$'.


\section {Measurements  and Quantum States Restrictions}

The essence of the state collapse problem is the 
  possibility of discrimination between 
pure and mixed states in quantum measurements, so
it's instructive to remind the classification of quantum states.
 First, there are the 
individual states - i.e.
 the quantum states in the individual events, in 
standard QM  they are 
the pure states - Dirac vectors $|\psi_l\rangle$ in Hilbert space $\cal H$;
see also chap.4  (Primas, 1981).
The   statistical states  describe the  properties
of quantum ensembles, and are represented by  the 
normalized, positive operators of trace $1$
- density matrixes  $\rho$ on $\cal {H}$.
 If   $|\psi_l \rangle$ composition  is known  for 
the given ensemble, it can be described in more detail by
 the   ensemble state (gemenge)    
 represented by the table $W^e=\{ \psi_l; P_l\}$, 
where $P_l$ are the corresponding probabilities (Busch,1996).
Here and below under standard QM we understand Schr$\ddot{o}$dinger
 formalism  which postulates that the quantum states
are defined on complex Hilbert space
 introduced $ad\quad hoc$ (Jauch, 1968).     
The properties of states in
   Algebraic QM     will be considered in chap. 3.

We shall use the  simple  MS  model
which    includes the measured state  S and
 IGUS  $O$   storing the incoming S information.
For the simplicity the detector D is dropped in MS chain,
   the role of $O$ decoherence effects will be discussed below.   
S represented by two-dimensional state vector $\psi_s$,
whereas $O$ is described by three-dimensional 
%
  Hilbert space $\cal {H}_O$. Its
  basis  consists of 
  the orthogonal states $|O_{0,1,2}\rangle$,
 which are  the eigenstates of $Q_O$  
'internal pointer' observable with eigenvalues $q^O_{0,1,2}$, which 
for the convinience is taken to be equal to $0,1,2$ correspondingly.  
For the simplicity of calculations it assumed that $|O_{1,2}\rangle$ 
constitutes the basis of two dimensional susbspace $\cal{H}'_O$. 
Let us   consider    
 the measurement   of  S observable $\hat{S}_z$
for MS initial  pure state:
\begin {equation} 
     \Psi^{in}_{MS}=\psi_s |O_0\rangle= 
(a_1|s_1\rangle+a_2|s_2\rangle)|O_o\rangle \, , \label {AAB}
\end {equation}
 where $|s_{1,2}\rangle$ are $Q$ eigenstates with eigenvalues $q_{1,2}$.
%
In our model  S,$O$  interaction $\hat{H}_I$ starts at $t_0$
and finishes effectively at some finite $t_1$;  
for the suitable $H_I$    
ansatz Schr$\ddot{o}$dinger equation for $\Psi^{in}_{MS}$ 
would result in
 MS final  state $\rho^p_{MS}$ responding to the state vector:
\begin {equation}
   \Psi_{MS} = \sum a_i|s_i\rangle|O_i\rangle \, .
                                   \label {AA2}
\end {equation} 
It turns out that 
$$
 \bar{Q}_O=\bar{S}_z=|a_1|^2-|a_2|^2
$$
, so $O$ performs the unbiased $Q$ measurement (Von Neumann,1932).
To reveal  the studied effect, the measurement of  incoming
  S  mixture with the same $\bar{S}_z$ should be regarded also,
  its composition is described by the gemenge 
$W^s=\{|s_i\rangle,|a_i|^2\}$ to which correspond the density matrix:  
\begin {equation}
 \rho_s^m= \sum_i |a_i|^2|s_i \rangle \langle s_i|
                                                              \label {AA33}
\end {equation}
%
Concerning  the information recognition by IGUS $O$,
 at this stage 
 we  assume arbitrarily  that $|O_{1,2,0}\rangle$ eigenstates
corresponds to 
the certain information patterns  $J^O_l$
percepted by $O$ as $Q_O$ values $q^O_{l}$.
Below  more arguments in favor of this assumption
will be presented, however it's worth to notice here that
in this case $q^O_i$ are $O$ real properties,
so this is quite natural assumption  (Busch, 1996).
 As the example of the simple $O$
 toy-model  can be regarded the hydrogen-like atom $A_H$
 for which $O_0$ is  its ground state
 and $O_i$ are the metastable  levels excited by $|s_i\rangle$,
for $|s_i\rangle $ superposition it results
  into the final S - $A_H$ entangled state.

%


The pure state  $\Psi_{MS}$ of (\ref {AA2}) 
   describes MS state relative to external observer $O'$,
whereas from $O$ 'point of view' $\Psi_{MS}$ describes the
simultaneous superposition (coexistence) of two
contradictory information patterns (impressions):
 $Q_O=q^O_1$ and $Q_O=q^O_2$ percepted by $O$
 simultaneously (Wigner,1961).
However, it's well known that experimentally   
the macroscopic   $O$ would observe at random one of $Q_O$ values $q^O_{1,2}$
in any individual event. It means 
that  S final state is $|s_1\rangle$ or $|s_2\rangle$ for $O$,  and
S state collapse occurs. 
   S final  state  is described by $\rho^m_s$
which stipulated by the gemenge 
$W^{MS}_{mix}=\{ |s_i\rangle|O_i\rangle, |a_i|^2 \}$.
In accordance with it one
 can  ascribe to MS the mixed state:
\begin {equation}
 \rho_{MS}^m= \sum_i |a_i|^2|s_i \rangle \langle s_i||O_i \rangle \langle O_i|
                                                              \label {AA3}
\end {equation}
relative to $O'$;
it    differs principally from $\rho^p_{MS}$ of (\ref {AA2}). 
  This   discrepancy  describes the phenomenon of state collapse,
     it illustrates also  the famous  Wigner 
  'Friend  Paradox' for $O, O'$ (Wigner,1961).
 
 We shall regard here   the formalism which temptatively can  unify 
this two    MS descriptions 'from outside' by $O'$
and 'from inside' by $O$. The first clue prompts the comparison
of $O$ average response to S pure and mixed states. Let's
consider for the simplicity only the incoming  $\psi_s$
with $r_s=\frac{a_1}{a_2}$ real, and denote them as $\psi_s^r$, 
 we introduce also  the z-symmetric state $\psi^s_s$
 for which $r_s=1$.
 For incoming  $\psi^r_s$, one obtains 
 $\bar {S}_x=2a_1a_2,\,\bar{S}_y=0$; 
 for $s_i$ mixture $W^s$ of (\ref {AA33}) $\bar{S}_x=0, \bar{S}_y=0$, so the 
  rate of S state purity $\beta_p$  
 is characterized by $|\bar {S}_x|$.
Meanwhile, beside $\bar{Q}_O=\bar{S}_z$, for any other $O$ observable
 $Q'_O  \ne F(Q_O) $ the final $\bar{Q}'_O=0$ independently of the initial
S state. Hence 
 the information about the rate of state purity characterized by $\beta_p$
 doesn't transferred to $O$ at all, at least via $O$ expectation values.
That's not quite surprising, because $S_z, S_x$ doesn't commute,
and so from the uncertainty relations follows that the precise $S_z$
meausrement would spoil $S_x$ measurement.
The analogous results 
 can be obtained for an arbitrary $a_{1,2}$,
but in place of $S_x$ some linear combination of $S_x, S_y$
should be used.
In Information Theory framework  MS and any other
  measurement set-up can be regarded formally as the information
channel transferring the information from S to $O$.
   Despite that the conclusive picture can be obtained only
from the analysis of individual  events, 
this calculations  indicates that
 the information transfer restrictions
 can be important  in    Quantum Measurement
theory.

From the formal point of view  the  measurement 
 of an arbitrary  system $S'$ is the 
mapping of $S'$ states set $N_S$ to  the given IGUS $O^I$ states set $N_O$
(Svozil,1993).   
   $O^I$ can be considered  formally as
 the subsystem of the large system $S_T=S'+O^I$ with the states set $N_T$
(Mittelstaedt,1998).
  In this approach - 'measurement from inside', $N_O$ is $N_T$ subset and 
  the inference map $M_T$ of $S_T$ states   to  $N_O$  defines
$O^I$ state  $R_O$ called   $S_T$ restricted  state.
 The important property of
 $S_T \rightarrow O^I$  inference map is formulated by
 Breuer theorem: if for two arbitrary $S_T$ 
states $\Gamma_{S},\Gamma'_{S}$ 
their restricted  states $R_O, R'_O$ coincides, then for $O^I$ 
observer this $S_T$ 
states are indistinguishable (Breuer,1996).
Under simple assumptions about $S_T,O^I$  at least several such  $S_T$  
states should exist.
  In classical case the origin of this  effect   is obvious:
  $O^I$ has less degrees of freedom  than $S_T$ and 
 can't discriminate all possible $S_T$ states, and because of it
some number of $S_T$ indistinguishable states 
 always should exist for any classical $S_T$
 and $O^I$ (Svozil, 1993).
 In quantum case  
  the  observables noncommutativity and nonlocality introduces
some novel features regarded below.
Despite that $R_O$ are incomplete $S_T$ states,
 they are the real physical states
for $O^I$ observer - 'the states in their own right', as Breuer 
characterizes them. 

 The obtained
$S'$,$O^I$,$S_T$ relations are applicable to our MS model which also can be
treated as  MS measurement from inside. 
  Breuer's results leaves the considerable freedom for the choice of
the inference map $M_T(S_T \to O^I)$, and
 doesn't permit
to derive the restriction ansatz for $S_T$ individual states  
directly.
It can be shown   that
 $R^{st}_O$ -  the restricted statistical state
 coincides with the  partial trace of $S_T$ state over $S'$
 degrees of freedom; 
  for MS   state  (\ref {AA2}) it is equal to:
\begin {equation} 
   R^{st}_O=Tr_s  {\rho}^p_{MS}=\sum |a_i|^2|O_i\rangle\langle O_i|
      \label {AA39}
\end {equation}
%
%
%
From that for   MS mixture $\rho^{m} _{MS}$
of (\ref {AA3}) 
 the  corresponding restricted 
 state is the same $R^{mix,st}_O=R^{st}_O$.
This equality doesn't mean  the
 collapse of MS pure state $\Psi_{MS}$,
 because  the collapse appearance
   should be   verified also  for  MS individual
  states.
For the pure case MS individual state is  $\Psi_{MS}$ of (\ref{AA2}), 
 for  the incoming S  mixture  - Gemenge $W^s$ of (\ref {AA33})
 MS state   also differs from event to event:
\begin {equation}
\varsigma^{MS}(n)=|O_l\rangle \langle O_l|| s_l\rangle\langle s_l|
  \label {A44}
\end {equation}
where the random $l(n)$ frequency is stipulated by the
 probabilistic distribution $P_l=|a_l|^2$, so 
in any event $\varsigma^{MS}(n)$  differs from
 MS  state (\ref{AA2}).
It was proposed phenomenologically that for an arbitrary system $S_T$
its restricted individual state   is  also equal to the
partial trace of $S_T$ individual state over $S'$ (Breuer,1996), 
 so for MS pure state:  
\begin {equation} 
   R_O=Tr_s  {\rho}^p_{MS}=\sum |a_i|^2|O_i\rangle\langle O_i|
      \label {AA4}
\end {equation}
Obviously such ansatz excludes beforehand any  
kind of probabilistic $R_O$ behavior.
Thereon  in  any event $n$ for MS Gemenge of (\ref {AA33})
its restricted individual  state  
 $\varsigma^O(n)=|O_l\rangle \langle O_l|$
   differs   from $R_O$  
(below  also $\varsigma^O_i =|O_i\rangle \langle O_i|$ notation is  used).
 It follows then, that   for the
restricted individual states the main condition of Breuer  Theorem
is violated. From that Breuer concluded
that  $O$ can differ pure/mixed states 'from inside'
in the individual events, therefore  this  
formalism doesn't result in 
 the state collapse  
 (Breuer, 1996). However,  the formal
difference of two  restricted states doesn't mean automatically
that this states are physically different.
That's the necessary but not sufficient condition,
in general  it's necessary also
that  some MS parameters measurable  by   $O$  
 are different for that states.
 Otherwise it can turn out that this  states are equivalent,
and $O$ can't detect their difference
(Mittelstaedt,1998). For $\varsigma^O_i$ 
the only parameter i.e. observable value
  which characterizes it in an individual events
is $q^O_i$. In Breuer paper $R_O$ parameters available for $O$
      weren't  calculated, without it
  the conclusion about $R_O,\varsigma^O_i$
 discrimination  seems preposterous;
the  discussion of this point will be continued in   chap. 4.
 Below by the slight abuse of terminology,
  MS state restriction to O is called $O$ restricted state. 
%


As was shown above the difference between 
 the initial S pure state and
$|s_i\rangle $ mixture with the same $\bar{S}_z$ reflected by 
$\bar{S}_{x,y}$ values. As the  example, for  $\psi^s_s$ with
 $a_{1,2}=\frac{1}{\sqrt{2}}$ for pure and mixed state one obtains
 $\bar {S}_x=1,0$   correspondingly.    
For MS   states  the same difference reflected by 
  MS interference term (IT) observable:
\begin {equation}
   B=|O_1\rangle \langle O_2||s_1\rangle \langle s_2|+j.c.
    \label {AA5}
\end {equation}
which characterizes $O$,S quantum correlations.   
Being measured by external $O'$ on S,$O$,
 it gives $\bar{B}=0$ for the mixed MS state  of (\ref {AA3}),
 but  $\bar{B}\neq 0$  for the pure MS states (\ref{AA2}).
For example, for the incoming $\psi^s_s$ one obtains $\bar{B}=1$. 
 However,  $B$ value  can't be measured by $O$ 
 'from inside', so $O$,S correlations are 
 unavailable for $O$ directly. The 
principal possibility for $O'$ to  measure $B$
and send the information to $O$ doesn't change 
the situation, because such $O'$ measurement is incompatible
with $S_z$  measurement by $O$.
 On the whole IT observables are nonlocal
and constitutes the special class of observables $\{B^{MS}\}$.
%
Note that $\{B^{MS}\}$  observabilty excludes
the Ignorance Interpretation (II) of pure states,
which assumes, in particular, that  MS  pure state
  $\Psi_{MS}$ is equivalent to $|s_i\rangle|O_i\rangle$
gemenge (Busch,1996).
As the example, for the symmetric $\Psi_{MS}$ of (\ref{AA2}) 
 with $a_{1,2}=\frac{1}{\sqrt{2}}$ II  
 claims that  $Q_O$ value in any event is sharp and
is equal either to $q^O_1$ or $q^O_2$ with the same probability.
Meanwhile  in this state
$B$ has  the eigenvalue $\tilde {b}=1$
  which means that   $Q_O$ value $\tilde{q}^O$
 is principally uncertain for $O'$:
 $q_1^O \le \tilde {q}^O \le q^O_2$, and so II is incorrect for $O'$.  
However such reasoning fails for $O$ observer, because
$B$ value is unobservable for him together with  other
 $B^{MS}$ observables.
   Hence the 'subjective' II,
 in which $Q_O$ value can be sharp for $O$,
being  simultaneously uncertain for $O'$, can't be excluded
beforehand  in self-description approach. 

%
%
%

In the regarded approach
MS individual state  can be rewritten formally in doublet 
form $\Phi^B(n)=|\phi^D,\phi^I \gg$ where $\phi^D=\rho_{MS}$
is the dynamical
state component i.e. MS state for $O'$,   the information component
 $\phi^I$ describes $O$ subjective information, i.e.
$O$ restricted state  in the given event $n$.
 In Breuer theory
 $\phi^I=R_O$ is just $\phi^D$  trace, however
in the alternative theory regarded here it  describes the independent 
$O$ degrees of freedom.
To agree with the
 quantum Schr$\rm\ddot{o}$dinger dynamics,
 any self-description theory should satisfy
 to the following operational conditions : \\
i) if an arbitrary system $S'$   doesn't interact with IGUS $O^I$,
 then for $O^I$ this  system evolves according to
 Schr$\rm\ddot{o}$dinger-Liouville  equation  (SLE)  \\
ii) If $S'$ interacts with $O^I$ and
 the measurement of some $S'$
observable occurs, then the  quantum dynamics can be violated for $O^I$, but
  as follows from  condition i),   
  in this case  $S',O^I$ evolution 
 for the external   observer  $O'$ should obey SLE.\\
To reconcile this conditions with the state collapse
at the phenomenological level we proposed recently 
 doublet state formalism (DSF) 
 (Mayburov,2001,2004). It's reviewed here briefly, because
 it has much in common with the
 description of measurements  in Algebraic  QM. 
For DSF state   
 $\Phi=|\phi^D, \phi^I\gg$ 
the dynamical component  is also
equal to QM density matrix  $\phi^D=\rho$ and obeys     SLE  :
\begin {equation}
  \frac {\partial \phi^D}{\partial  t} =[\phi^D,\hat {H}]    \label {AA8}
\end {equation}
  therefore  for MS
the initial $\phi^D=\rho^{in}_{MS}$ of (\ref{AAB}) evolves   
 at $t>t_1$ to $\phi^D(t)=\rho^p_{MS}$ of  (\ref{AA2}).
$\phi^I$ describes $O$ restricted state, so
   for $t \le t_0$,  the initial
 $\phi^I=|O_0\rangle \langle O_0|$. 
   After S measurement finished
 at $t>t_1$,  in DSF its $\phi^I$ outcome  is supposed to be stochastic:
 $\phi^I(n)=\phi^I_i$,  where
$\phi^I_i=|O_i\rangle \langle O_i|$;
 here $i(n)$ outcome frequencies are
 described by  the probabilistic distribution with $P_i=|a_i|^2$.
 Hence such doublet individual state $\Phi(n)$ 
can change from event to event,  and $\phi^I(n)$
is partly independent of $\phi^D$, being correlated with it only
 statistically. Consequently the ensembles of
  $O$ subjective states $\phi^I$  coincides for
the pure and mixed states with the same $|a_i|^2$, the conditions of
 Breuer theorem are fulfilled and  the subjective state
collapse can be observed by $O$.

Plainly in this theory 
the quantum  states for external $O'$ (and other observers)  also has
the same doublet form $\Phi'$. In the regarded situation
 $O'$ doesn't interact with MS  and so  $O'$ information $\phi^I$
doesn't change during S measurement. Consequently,
 MS  state evolution  for $O'$ described by
$\phi^D$, which obeys SLE. Because of it  MS state
collapse isn't observed by  $O'$ in agreement with  the conditions i, ii;
eventually this theory responds to the subjective II
regarded above.
  Witnessing QM Interpretation proposed by Kochen  
  is quite close to DSF but doesn't exploits the self-description approach
(Kochen,1985; Lahti,1990).   

%



%




In DSF $|O_i\rangle $ constitutes the  
 preferred basis (PB) in $\cal H_O$;
this problem called also the basis degeneracy   is well-known
 in  Quantum Measurement Theory  (Lahti,1990; Elby,1994). 
In its essence, the theory consistency demands that the final
 MS states decomposition should be unique,
 but this isn't the case for $\Psi_{MS}$ of (\ref {AA2}).
In DSF  PB problem acquires the additional aspects related
to the information recognition by $O$. 
The plausible explanation of PB appearance prompts $O$ decoherence - i.e.
$O$ interaction with environment E,
which is practically unavoidable in lab. conditions (Zurek,1982).
Such interaction  results in the final entangled S,$O$,E state
which decomposed on some orthogonal $O$ basis  $|O^E_i\rangle$
 (Guilini,1996). 
 Tuning the   $H_{O,E}$ interaction
parameters,   $|O^E_i\rangle$ basis can be made equivalent to 
$|O_i\rangle$ basis. 
For the initial MS state $\Psi^{in}_{MS}$ of
 (\ref {AAB}) it results in the final MS-E state :
\begin {equation}
    \Psi_{MS+E}=\sum a_i|s_i\rangle|O_i\rangle|E_i\rangle
                                             \label {DD1}
\end {equation}
where  $|E_i\rangle$ are final E  states.
It was proved that such triple decomposition is unique, even
if $|E_i\rangle$ aren't orthogonal (Elby,1994).
  This measurement 
scheme denoted as MS+E  will be considered below
together with  MS model.
 In other aspects the
 decoherence doesn't change our measurement model;  its
most important role is the unambiguous definition of $O$ PB.
In fact $\cal {H}_O$ symmetry is broken dynamically 
by $H_{O,E}$ interaction which makes majority of $O$ states unstable.
As will be shown   in Algebraic QM PB is defined by S,$O$ interactions
only and is independent of $O$ decoherence,
 however the  account of decoherence effects is necessary in any consistent   
measurement theory.



\section {   Quantum Measurements in Algebraic QM}

Now   the quantum  measurements and $O$ self-description
 for the finite quantum systems will be regarded
in  Algebraic QM formalism; analogously to DSF
it should also satisfy to the conditions i) - ii) formulated above.
  Besides the standard quantum effects,
Algebraic QM    describes successfully
the phase transitions and other nonperturbative phenomena 
which  standard QM fails to incorporate (Emch,1972). 
Consequently, there are the serious reasons to regard Algebraic QM
as the consistent generalization of standard QM.
Algebraic QM was applied extensively
 to the superselection models of quantum measurements,
in which the detector D or environment E  are regarded as the infinite systems
with $ m,V \rightarrow \infty $ (Primas,1990; Guilini,1996).
The algebraic formalism of nonperturbative QFT  
was used also in  the study of measurement 
dynamics in some realistic  systems (Mayburov,1998; Blanchard,2003).
%
%

What is the fundamental QM entity -
the Hilbert space of states or the algebra of observables
is  the problem dusputed since  the time of QM formalism construction 
 (Von Neuman, 1932; Emch,1972).
In standard QM the fundamental structure is the 
 Hilbert space  of states $\cal H$
 on which an observables - Hermitian operators
( or POV) are defined (Busch,1996).
 However, it was found that for some nonperturbative 
systems  the structure of states set   differs principally
 from  $\cal{H}$, and the axiomatics of standard QM
 becomes preposterous (Bratelli,1981). In distinction, 
   the fundamental structure  of Algebraic QM
 is   Segal algebra $\cal{U}$ of observables
which incorporates the  main properties of any  regarded system $S_f$
 (Emch, 1972).
From the mathematical point of view 
the duality between the operators algebra and the states set
is more natural, than the states set priority
postulated in standard QM (Bratelli, 1981).
In its framework  it's more   convenient technically   to  deal with 
 $C^*$-algebra $\cal{C}$  for which $\cal{U}$ 
constitute  the subset.
 $\cal C$ and $\cal U$ elements in Algebraic QM
  are the linear operators for  which the sum $A+B$ and
   product $A \cdot B$  defined.
  Roughly speaking, $\cal{C}$ is the complex algebra 
  for which $\cal{U}$ is the subset 
 of its real (Hermitian) elements.
 For any system $\cal{C}$, $\cal{U}$ are in the unambiguous
 correspondence: $\cal{C} \leftrightarrow \cal{U}$,
and below their use is equivalent in this sense.
   $S_f$   states set $\Omega$ defined by
 $\cal{U}$    via the notorious
 GNS  construction; in its framework
  $\Omega$ is the  vector space dual to
 the corresponding  $\cal C$ (Bratelli,1981).
 Such states are
called here the algebraic states $\varphi $ and
are  the normalized, positive, 
linear functionals on $\cal {U} $: for any observable $A \in \cal{U}$,
  $ \, \forall \varphi \in \Omega;$, its expectation value
$\bar{A}=\langle \varphi;A\rangle$. 
Eventually $\varphi$ states are the analog of QM density matrixes $\rho$.
%
%
The  pure states -  i.e.   $\Omega$  extremal points are 
  regarded  as the algebraic individual states
 (AIS) $\xi$, their set  which is a convex shell
denoted $\Omega^p$;
 the further consideration of AIS properties
is given below in chap.4 (Emch,1972; Primas,1990).
%
All  other   elements $\varphi \in \Omega$ are treated
as the algebraic mixed states, despite that their physical
meaning isn't settled finally (Primas, 1981).    
Independently of it any algebraic  state $\varphi$
can be constructed operationally as $\xi_i$ ensemble,
  the   ensemble states $W^A$ are
  defined analogously to  QM ansatz given in chap.2.
Here only  a finite-dimensional $S_f$ will be considered,
in this case both regarded QM formalisms are principally equivalent, 
but Algebraic QM is more convinient for our problems,
because in its framework  the correspondence between states and 
observables is more straightforward.
In this case $\varphi$ states set $\Omega$ is isomorphic to $S_f$ set of
 QM density matrixes $\rho$.


Concerning  the application of Algebraic QM to the measurements,
 in particular  for our MS scheme, 
 $O$ self-descripiton restrictions can be formulated 
 as the restrictions
on the set of MS observables  available for  $O$. 
The  situations 
in which for some system $S_f$ only some restricted 
   linear subspace $\cal M_R$  or  subalgebra
 $\cal{U}_R$ of   algebra $\cal U$
is available for the observation
were  extensively studied  in (Emch, 1972).
 In this case the restricted  algebraic states $\varphi_R$
can be constructed
 starting from  the expectation values of $A_R \in \cal{U}_R$:
\begin {equation}
    \bar{A}_R=\langle \varphi;A_R\rangle=\langle \varphi_R;A_R\rangle
  \label {CC12}
\end {equation}
 $\varphi_R$ doesn't depend on any $A' \notin \cal{U}_R$,
thereby $\forall \varphi_R,\,\langle \varphi_R;A'\rangle=0$.
%
%
In Algebraic QM  any classical system $S^c$  
described by the commutative system of observables $\{A^c_i\}$
which constitute the associative Segal algebra $\cal {U}^C$.
Conversely any  associative Segal  (sub)algebra $\cal{U}'$
is isomorphic to the algebra $\cal {U}^C$    which
 describes some $S^c$, its  $\varphi^a$ states set
  $\Omega^a$ is isomorphic to
  the set $\Omega ^c$ of the  classical statistical
$S^c$ states $\varphi^c$ (Segal.1947).
The corresponding AIS   - i.e. the pure states   
corresponds to the classical
individual  states $\xi^c_i$ - points in $S^c$ phase space.
For the  self-description the most important is the  case when
  $\cal{U}'$ is elementary, i.e. includes only $I$ - unit operator
 and  the single $A \ne I$ (it's also called A-subsystem). 
   Then $\xi^c_i=\delta(q^A-q^A_i)$ corresponds to    
 $A$ eigenvalues $q^A_i$ spectra.
 Consequently, even if quantum $S_f$ is
 described by nonassociative $\cal{U}$,
 it includes the subalgebras $\cal{U}'\in U$
   for which       the restricted states are classical.
Hence  in Algebraic QM $S_f$  state can be described
formally as the multiplet, each member of
which is the state defined on the particular (sub)algebra
 analogously to DSF state regarded above.
%
%

For the classical  system $S^c_T=S'+O^c$ 
 described by some $\cal{U}^C$, the self-description restrictions for
IGUS  $O^c$   are simple and straightforward -
the restricted $S^c_T$ states depend only on those
 $S^c_T$ coordinates $\{x^O_j\}$,
 which are $O$ internal degrees of freedom (Breuer,1996). 
They constitute the subalgebra $\cal{U}^C_R \in \cal{U}^C$.  In practice
 $O^c$ effective subalgebra $\cal{U}^C_O \in \cal {U}^C_R $ which
really defines the measurements 
 can be even smaller, because
some $x^O_j$ can be uninvolved directly into the measurement process.  
From the analogous reasons
   for  the quantum localized  IGUS $O$ 
 its subalgebra $\cal{U}_R$ of MS algebra $\cal U$ also should
include only $O$ internal (local) observables. 
It corresponds   to the locality principle  acknowledged
in Quantum Physics. Really $O$ perception of
 any other MS observable $A_{MS} \notin \cal U_R$
 involves  the instant  measurement of S which can be miles away
from $O$ at that time, the example is IT $B$ of (\ref {AA5}).
Any effective $O$ subalgebra  $\cal{U}_O \in \cal{U}_R$,
   their  states  sets are denoted $\Omega_O, \Omega_R$ correspondingly.   
Concerning with $O$ individual states 
the  main assumption of our theory is as follows:
 given  the subalgebra $\cal{U}_A \in \cal{U}$, 
  in any individual event $n$ an arbitrary  MS state $\xi^{MS}(n)$ 
 induces  some    restricted AIS $\xi^A(n)$ spanned on $\cal {U}_A$.
This hypothesis seems to us quite natural  and  
 below the additional arguments in its favor will be presented
for the particular $\cal{U}_A$.

%
%
  MS is described by
 $ \cal{U}$ Segal algebra  of MS  observables  which defines
  $\varphi^{MS}$ set  $\Omega$. In case of $O$ decoherence
  $MS+E$ involves $\cal {U}_{MS,E}$ algebra correspondingly.
 $O$  subalgebra  $\cal{U}_R$  includes $I$ and all $O$ 
internal observables, so it means that $\Omega_R$
 is isomorphic to $O$  statistical states $\rho^O$ set $\cal V_O$. 
 Then $O$ AIS set $\Omega^p_R$ is isomorphic to $\cal{H}_O$, and 
 any $O$  AIS $\xi^R_i $ should correspond to some  
 state vector $ |O^R_i \rangle \in \cal H_O$.
We don't study here $\xi^R$  states in detail, because the effective
$O$ subalgebra  $ \cal U_O $ differs from $\cal U_R$,
and our aim is to calculate $O$ states defined on it.
%
To define   $\cal U_O$,
 remind that
   in the used MS dynamics of S measurement   for any final
$\Psi_{MS}$ 
  only  $\bar{Q}_O \ne 0$ in MS final state, $\bar{Q}_O=\bar Q$; 
for any other $Q'_O \neq F(Q_O)$   one obtains  $\bar{Q}'_O=0$.
 Therefore for  any corresponding AIS $\xi^O \in  \Omega^p_O$
 it  follows that  
 $ \langle  \xi^O; Q'_O \rangle=0$.
Any $\varphi^O \in \Omega_O$ is the convex state of $\{\xi^O \}$, hence
$\langle \varphi^O,Q'_O\rangle=0$.
%
%
%
It means that  
  $O$ effective    subalgebra $\cal{U}_O$ is equal to
the elementary $\cal U^I_R$, which
 includes only  $Q_O$ and $I$.
Really, only in this case  
$ \langle   \varphi'; Q'_O \rangle=0$
 for all $ \varphi' \in \Omega^I_R$ defined on  $\cal U^I_R$;
 each $\varphi'$ corresponds to $\varphi^O$ with the same $\bar{Q}_O$
and vice versa;
 so  $\varphi^O$ set $\Omega_O$  is isomorphic to $\Omega^I_R$ 
(Segal, 1947).
 There is no other $\cal U_R$ subalgebras with such properties, and
  that settles $\cal U_O $ finally as the classical algebra of
single 'pointer' observable $Q_O$. Consequently 
such $\cal U_O$ defines also PB for $O$ states unambiguosly,
and so $O$-E decoherence only can only duplicate it;
  more detailed consideration of the decoherence effects  is given below.

%

 \section {Algebraic QM Restrictions and Self-description} 


Now   the relation between  the  pure  and individual states
 is regarded in more detail, because   it will be used
 below in the construction
 of  restricted $O$ states in Algebraic QM -  AIS.
 Despite that mathematically
this question is quite plain, it's 
  considered here in most detailed way,  because in the discussion
of QM foundations it became 
the source of many confusions   (Bene,2000; Breuer, 1996).
In this consideration of individual  quantum states
 i.e. the states distinguishable in the individual events
   we shall follow the approach  of (Primas, 1990).
As was mentioned in chap. 2,
the  physically different states of any kind can be  operationally
discriminated by the particular experiment
which puts in correspondence  to this states some
measured parameters values.
For QM statistical states  it can be the parameters of  
 experimental distributions  which can be expressed via
the expectation values of some observables. 
 For the finite-dimensional nondegenerate $S_f$ any pure state $\psi_{a}$
is   the eigenstate of some  observable $A$  
with the eigenvalue $q^{A}_{a}$; in this case
 $q^{A}_a$ corresponds to some real $S_f$ property (Busch, 1996).
 It permites $\psi_a$ discrimination from 
any  other pure state $\psi_b$ in a single event
by the objective (non-collapse) $A$ measurement,
  hence  any $\psi_a$ is the individual state. 
There is no other classes of states which satisfies to this conditions, 
in particular, the convex states can't be an individual states.
To illustrate it, let's regard as
 the  example  S state $\psi_s$ of (\ref {AAB}),
 for an arbitrary $a_i$ it is the eigenstate of some operator
$\vec{S}\vec{n}$, where $\vec{n}$ direction defined by $a_i$.
Now let's  consider   the convex state
$ \phi_z=\sum w_i|s_i\rangle \langle s_i|$ with $w_i \ge 0$, in this
     framework it has no real properties. Hence it's
impossible to prove that $\phi_z$ is  the superposition
of states with $s_z= \pm 1$, alike the real property 
 $s_x=1$ demonstrates it for $S_x$ eigenstate $\psi_s^s$ with
 $a_{1,2}=\frac{1}{\sqrt{2}}$.
$\phi_z$ is smeared on $s_z$ axe, i.e. $-1 \le s_z \le 1$, but the
 same inequality is true for $|s_{1,2}\rangle$ states also. Consequiently
 by no means
 $\phi$ can be discriminated from $|s_{1,2}\rangle$ 
in the individual events, and because of it
 $\phi$ can't be the feasible individual state. 
In this framework the ansatz for 
the individual states  is defined by the operational conditions only,  
and hence will be  principally the same in Algebraic QM -
i.e. AIS should be also  the pure states. For the restricted  algebraic
states this  ansatz doesn't change principally,  as shown below
    for $\cal U_O$ its derivation is even more simple and straightforward. 

As follows from Segal theorem,
for the obtained $\cal U_O$ subalgebra
     the  algebraic $O$ 
 states $\varphi^O\in \Omega_O$   are isomorphic to the classical
 $q^O_i$ probabilistic  distributions;
meanwhile  $\xi^O$ -  $\Omega_O$ extremal points
 are the positive states:
\begin {eqnarray}
            \xi^O_i= \delta(q^O-q^O_i) \label {AA66}
\end {eqnarray}
 which corresponds to the classical pointlike  states.
%
  In particular, such $\xi^O_i$  appears in 
$|s_i\rangle$  measurement  by $O$
 as the restriction of MS final
state $\xi^{MS}_i \sim |s_i\rangle|O_i\rangle$.
%
 $\xi^O_{i}$  are $O$ individual states, their 
distinction  can be revealed in the single event from the
the difference of $Q_O$ eigenvalues $q^O_i$. The
existence of such restricted $O$ states at least for some MS states
 restriction is  important for our formalism, because it  
permits to analyze  any other  restricted states (if they exist)
 by the comparison with $\xi^O_i$. Following the   
consideration  of chap. 2, we concede that $O$ percepts this states as
the information pattern $J^O_i=q^O_i$.

%
 
 The incoming S mixture - $|s_i\rangle$ gemenge  $W^s$ results in
 MS algebraic final state 
  $\varphi_{mix}$ which is equivalent to $ \rho^m_{MS}$ of  (\ref {AA3}); 
%
%
  $O$ restricted state  $\varphi^O_{mix}$ 
is  defined from the relation for $\bar{Q}^n_O$, in particular :
$$
    \bar{Q}_O=\langle\varphi^O_{mix} ;Q_O\rangle=
    \langle\varphi_{mix};Q_O\rangle=\sum |a_i|^2 q^O_i
$$
which results in the solution  $\varphi^O_{mix}=\sum  |a_i|^2\varphi^O_i$,
where $\varphi^O_i=\xi^O_i$    of (\ref {AA66}).
From the  correspondence
of MS state $\xi^{MS}_i$ and $O$ state $\xi^O_i$ in the individual events,
the restricted algebraic  state $\varphi^O_{mix}$
 represents in this case
the statistical mixture of AIS $\xi^O_i$ 
 described by $O$  ensemble state 
\begin {equation}
  W^O_{mix}=\{ \xi^O_i;\,P_i=|a_i|^2;\,i=1,2\} \label {AA55}
\end {equation}
%
 If MS final state is  the pure state $\xi^{MS}$
which corresponds to $\Psi_{MS}$ of (\ref {AA2}),
 then MS  algebraic state is
$\varphi^{MS}=\xi^{MS}$, and  it results 
 in  the same $\bar{Q}^n_O$ value as the regarded mixture.
Therefore its $O$ restricted algebraic state
coincides with the mixed one: $\varphi^O=\varphi^O_{mix}$ 
for the same $|a_i|$.

To illustrate the  derivation  of MS restriction to $O$ 
for  individual states,
let's consider it for $\xi^{MS}_s$ which is equivalent to $\Psi_{MS}$
 of ({\ref{AA2}) for  z-symmetric initial state $\psi^s_s$ with
 $a_{1,2}=\frac{1}{\sqrt{2}}$.
 Let's express   the properties  which
MS  restricted  state - $\xi^O_s$ should possess via  
 its relations with  the  observables
  $Q_O$ and $B$ of (\ref {AA5}), their values (in general uncertain)  
 are denoted as $\tilde{q}^O,\tilde{b}$.
In $O'$ RF $\xi^{MS}_s$ is $B$ eigenstate with the eigenvalue $\tilde {b}=1$ 
which is called   IT property;
 $Q_O$ obeys the inequality:
 $q^O_1 \le \tilde {q}^O \le q^O_2$, it is called the spectral property.
Taken together this properties indicates that for $O'$ $\tilde{q}^O$ is 
 located within the interval $[q^O_1,q^O_2]$, and  so is
principally uncertain inside it,
as $\tilde b=1$ value evidences.
  $\xi^O_s$ is defined on
 $\cal {U}_O$ $=\{I,Q_O\}$,
therefore the spectral property
 also holds for $O$. Now because $B \notin \cal{U}_O$
and so is unavailable for $O$, for $\xi^O_s$
IT property can be dropped.
  Without it the spectral property alone
  means  that $\tilde{q}^O$ is   localized in $[q^O_1,q^O_2]$ interval,
 but $\tilde{q}^O$  can be either uncertain or sharp  in its limits.
 Meanwhile $\xi^O_{1,2}$ states with  the sharp $\tilde{q}^O=q^O_{1,2}$
  possess   that only  property which MS restricted state should have.
Therefore the
 solution for $\xi^{MS}_s \rightarrow \xi^O_s$ restriction in the 
individual event $n$  can be formally written as:
 \begin {equation}
          \xi^O_s(n)=\xi^O_1.or.\xi^O_2   \label {B3} 
\end {equation} 
which means that in each event $\xi^O_{1}$ or $\xi^O_2$  appears at random.
%
%
To reproduce the correct expectation values $\bar{Q}^n_O$,
the corresponding $\xi^O_i$ probabilities are $P'_i=|a_i|^2$.
Then for any $\xi^{MS}$ there is always the probabilistic ensemble 
 $W^O=W^O_{mix}$  of (\ref {AA55}) which describes
 consistently the properties of MS  restricted state. 
%
One must also show that no 
 alternative nonprobabilistic solution $\xi^O_f$ 
for MS restriction exists.
  $\xi^O_f$ should have   
 correct the expectation values $\bar{Q}^n_O$,
the only nonstochastic
state $\xi^O_f \in \Omega_O$ which obeys this conditions is
$\xi^O_f=\sum|a_i|^2\xi^O_i$, the analog of Breuer state $R_O$.
%
Such $\xi^O_f$ should be a nonlocalized state on $Q_O$ for which 
 $q^O_1 \le \tilde{q}^O \le q^O_2$, i.e. $\tilde{q}^O$ is uncertain.
 As was explained above, in that case
some observable  $B'\in \cal U_O$ should exist
  which eigenvalue or an expectation value  $\bar{B}'$ for $\xi^O_f$
 would reveal its difference from $\xi^O_i$ states and
 demonstrate   $\tilde{q}^O$  genuine uncertainty,
i.e the simultaneous presence in several   $Q_O$ points.
Yet $\cal{U}_O$ doesn't include any   other observables beside $Q_O$,
thereon $\xi^O_f$ isn't feasible as $O$ individual state 
and  $\xi^O_{s}$ is the only suitable solution. 

The same arguments are applicable for an arbitrary $a_i$,
 the only difference  is that another IT $B^a \in \{B^{MS}\}$ is
 involved in the derivation; eventually the same unique solution $\xi^O_s$ of 
  (\ref {AA55}) exists, and $W^O$ with the 
given $|a_i|^2$  describes its ensemble.
 In general 
he nonexistence of any $O$ individual
states $\xi^O_f \ne \xi^O_i$ directly follows from
$\cal U_O$ elementarity, because as was shown, all individual 
states should be the eigenstates of some observables and in this
case it can be only $Q_O$.
%
Consequently in Algebraic QM  
$\xi^{MS}\rightarrow \xi^O$ inference map is stochastic 
and results in  the  subjective state collapse  observed by $O$.
 MS AIS can be expressed as the doublet
 $\xi^{MS}$, $\xi^O$, which corresponds to DSF dynamical and 
information components. Correspondingly MS evolution operator $\hat{Z}(t)$
is also the multiplet (doublet) which includes the unitary component
for $\xi^{MS}$ and nonunitary stochastic one for $\xi^O$. 
 In the regarded situation
 $O'$ doesn't interact with MS,  and so
analogously to DSF formalism  $O'$ information 
doesn't change during S measurement.
 Because of it  MS state
collapse isn't observed by  $O'$ in agreement with  the conditions i, ii
of chap.2.  Note that the same results can be obtained if in place 
of $\cal U_O$
we shall exploit MS restriction to $\cal U_R$ - the subalgebra 
of all $O$ observables.

For illustration 
let's regard the opposite hypothesis: namely that $O$ 
 percepts pure MS state $\xi^{MS}$ as $Q_o$ 'pointer superposition'
and it differs from $O$ perception of regarded MS mixture;
the analogous situations were widely discussed 
(Von Neumann,1934; Wigner,1961).
 If to assume that $O$ perception doesn't violate QM laws,
   in particular, that all $O$ observations are related to some 
  $O$ observables, then
 it should be some $O$ observable $G_O$ -
'the number of $Q_O$ peaks'  for which $\bar{G}_O=1,2$ for MS
mixture and pure state correspondingly. But 
such $O$ observable doesn't exist not only for $\cal U_O$, but
  even for $\cal U_R$ subalgebra. 

%


%

 Note that MS symmetric individual states $\xi^{MS}_s$ possess the
reflection symmetry $Q_O \to  - Q_O$, but no
  $O$ states $\xi^O_i$  have that property.
In $C^*$algebras formalism of Quantum Field Theory
 such symmetry reduction results in
 the phenomena of Spontaneous Symmetry Breaking; in particular, it leads to 
the randomness of outcomes for some models of  measurements 
in the collective
systems (Guilini,1996; Mayburov,1998).
The  self-description approach  permits to extend
such randomness mechanism on the finite systems
measurements. Its most outstanding feature is the simultaneous
coexistence of two individual states $\xi^{MS}, \xi^O_i$ which
 seems contradictive.
From  the mathematical point of view the
 Algebraic QM  contains the generic structure -
the restricted AIS set $\Omega^I_R$ defined on  the
elementary subalgebra
 $\cal {U}^I_R \in \cal U$ 
(in our model $ \cal {U}^I_R=\cal{U}_O$).
  $\Omega^I_R$ extremal points are treated as the individual
 states.
Hence  in this theory
the quantum state reduction results from the reduction of  a system 
algebra to its associative subalgebra. Eventually MS, $O$ states
are defined on different algebras which corresponds
to observations in $O,O'$ RFs,
 and so they are unitarily nonequivalent. 

If to analyze obtained results  in the Information-Theoretical framework,
 remind that
 the difference between the pure and
mixed MS states reflected by the expectation values
$B$  of (\ref {AA5}) and other IT observables.
Therefore $O$ possible observation of S pure/mixed
 states difference would mean
that $O$ can acquire the information on   $B$  expectation value.
 But $B \notin \cal {U}_R$ 
because of it S,$O$ IT correlations
 are unobservable from inside by $O$ (Mittelstaedt,1998).
Therefore the information  which describes the difference of the pure
and mixed S states principally can't be transferred in such MS scheme
  from $S$ to IGUS $O$ 
for the consequent processing.
It corresponds with Wigner conclusion that the perception by $O$
of the superposition
of two contradictive impressions is nonsense
and should be excluded in the consistent theory (Wigner, 1961).
 Our  calculations can be regarded as  the kind of 'no-go'
theorem which prohibit   such superpositions observations.
In this context the obtained $\xi^O_s$ state describes
      the upper limit of S information available to $O$. 
Note that the formal  impossibility  of pure/mixed states discrimination  
for the restricted  set of observables
was considered  earlier (Busch,1996),
but its applicability to
 the feasible measuring schemes wasn't  proved.

 In practice
it's possible that $O$ effective subalgebra is larger than $\cal U_O$,
 but this case
 demands more complicated calculations which we plan to present
in the forcoming paper. In Algebraic QM 
 the only important condition for the classicality appearance
 is $\cal{U}_O$ associativity but it 
  feasible, in principle, also  for the complex IGUS structures.
As was shown 
the formalism of Algebraic QM  extracts PB
 $\xi^O_i \sim |O_i\rangle $ in $\cal {H}_O$ 
even without  the account of E decoherent interaction,
but only from  the uniqueness $O$ subalgebra $\cal {U}_O$
for a given S, $O$ interaction. 
 The $O$-E decoherence, in fact, only duplicates this effect
        resulting  in the optimal case in the same PB solution.  
 If to consider MS+E system 
and its $\cal {U}_{MS,E}$ algebra, the effective $O$ information subalgebra
will be the same $\cal {U}_O$ considered above. 
Therefore $O$ subalgebra and its states set
properties can't depend directly on the surrounding E properties,
and the final $O$ states structure is analogous to the obtained above.
%
%


As was shown in chap. 2 standard QM doesn't contain the unambiguous
      restriction  ansatz for individual states,
therefore some   additional assumptions are necessary
to introduce it, if one prefers to work inside its framework.
From our analysis it's sensible to assume that   
      for MS $\to O$ restriction   
   to any $\Psi_{MS}$ corresponds some $O$ individual state in any event.  
Undoubtly any pure state $\psi^O_j \in \cal H_O$ is $O$ individual
   state with some real property.
 Meanwhile  from the arguments given in the beginning of the
chapter, Breuer state $R_O$ of (\ref {AA4}) is the convex state and so
 can't be $O$ individual  state. But this is the only nonprobablistic  
solution which can dispatch the correct $Q_O, Q'_O$ expectation values.
Hence they can be restrored only by the random mixture of several
$\psi^O_j$ states, and the stochastic ensemble of
$|O_i\rangle$ states with the probability $P_i=|a_i|^2$ is the only solution.
Note that it
reproduces   the results of  our DSF ansatz which earlier was obtained 
phenomenlogically. Eventually under the simple minimal assumptions
the effect of state collapse can be obtained in the standard QM framework,
yet  it seems that Algebraic QM on whole proposes more consistent
 approach to the problem.

For the regarded simple MS model Algebraic QM formalism in many aspects
is analogous to Orthomodular Algebra of propositions or Quantum Logics
(Jauch, 1968; Emch, 1972). 
 Its possible application to the systems self-description 
and the measurement from inside  deserves detailed investigation,
here only few  notices are given.   
If one assumes that in this framework the
measurement restrictions  also are defined by the set of
$O$ local  observables -
i.e propositions available for $O$ observations, then 
 the results can be comprehended easily for the case when
only single observable $Q_O$ is available.
In this case the corresponding lattice $\cal L^O$ of propositions includes only 
$Q_O$ projectors   set - $\{P^O_i\}$,
 (together with $\ominus,I$) and hence is atomic
$\sigma$-continuous Boolean lattice. Then it follows that  
the set of restricted  $O$ individual states $\eta^O_i$ which
corresponds to this lattice
 is isomorphic
to classical $O$ states set $\{ \xi^O_i\}$ (Jauch, 1968). 
For any other state $\eta^O_R$ no proposition from $\cal L^O$
can be put in correspondence, and therefore such state  isn't feasible.
In general the search of alternative and 
   more simple formalism of self-description  
 deserves the thorough investigation. Really the only principal 
feature of Algebraic QM, which was used in our approach, 
is the strict correspondence of the restricted states and the set
of local observables, but it doesn't   demand that such set 
should constitute  Segal algebra.

Despite of the acknowledged achievements of Algebraic QM
   its foundations are still
discussed and aren't  settled finally. In particular, 
it's still unclear whether all the algebraic states $\varphi$ 
correspond to the physical states, this problem
discussed thoroughly in (Mayburov, 2005). 
This question is  important by itself 
and  can be essential for our formalism feasibility (Primas, 1983).
 We admitted  also
 that for  MS arbitrary  $\xi^{MS}$  some   $O$ restricted AIS responds
in any event.
 It agrees with the consideration of the restricted states  
 as the real physical states, 
 however, this assumption needs further clarification.



 \section { Discussion}

in this report the
 information-theoretical restrictions on the quantum measurements
were  studied  in the simple selfdescription  model of IGUS $O$.
Self-description theory shows that
by itself   $O$ inclusion  as the quantum object
into the measurement scheme doesn't result in the
 state collapse appearance (Breuer,1996).
  At the phenomenological level the appearance of the
 state collapse described by  DSF which exploits 
 the doublet states $\Phi$ ansatz,
where one of its components $\phi^I$ corresponds to $O$
 subjective information.
  Algebraic QM  
presents the additional arguments in favor of this
 collapse mechanism; its formalism reflects the strict correspondence
between the states and observables -  the principal distinction of
QM from Classical Waves theory.
 In our approach
 $O$  subjective state after the measurements
 is  defined on the set of $O$ internal observables  $\cal {U}_O$.
 Yet the observable $B$, which value characterizes the 
pure/mixed states difference, or more precisely the class of IT observables
 doesn't belongs to $\cal {U}_O$, hence $O$ internal state
can't react on this difference.
%
 In this paper Algebraic QM  was applied for 
the simple measurement model, but if this formalism universality
will be proved, it would mean that the proposed measurement
 theory follows from the established Quantum Physics realm (Emch,1972).
From the formal point of view the only novel feature of our approach 
is the use of Segal algebra for the individual $O$
 states $R_O(n)$ calculations.
Note also that no compelling mathematical arguments proves that  
 for the individual states $R_O$ 
 ansatz in standard QM framework
 should be chosen  the same QM formulae (\ref {AA4})   
  which used for  the statistical restricted states
$R^{st}_O$ (Lahti,1990).



Our  theory demonstrates that the probabilistic realization
 is generic and unavoidable for QM and without it QM supposedly can't
acquire any operational meaning. Wave-particle dualism
was always regarded as characteristic QM  feature but in our theory
it has straightforward  correspondence.
 Algebraic QM  approach stresses also the
 dual character of
quantum measurement : this is the interaction 
of studied S with IGUS $O$ and in the same
time the information acquisition and recognition by IGUS. 



On the whole
the decoherence of macroscopic objects is a very important effect,
under the realistic conditions the rate of E atoms interactions with 
macroscopic detector D is very high, and because of  it
 in a very short time $t_d$
S,D partial state $\rho_{SD}=Tr_E \rho_{SDE}$ becomes approximately equal
to the mixed one, as $\rho_p$ nondiagonal elements become
very small. This fact induced the claim
that the objective state collapse   can be completely explained by detector  
state decoherence without the observer's inclusion, but it
was proved to be incorrect (D'Espagnat,1990).
The further development of decoherence approach
which accounts the observer was  proposed
  in Zurek 'Existential interpretation (Zurek,1998).
 IGUS   $O$  regarded as the  quantum object and
 included in the measurement chain; 
  the memorization of input S signal  
occurs in several binary memory cells $|m^j_{1,2}\rangle$ (chain)
which are the analog of  the brain neurons. 
 $O$ memory state suffers  the decoherence from surrounding E 'atoms'
which results in the system state analogous to (\ref {DD1}).  
Under  practical  conditions, the decoherence time $t_d$ 
is also  small and for $t \gg t_d$
 S,$O$ partial state $\rho_p$ differs from the mixture
very little.  From that  Zurek concludes that $O$ 
percepts input pure S signal as the random measurement outcomes.
However the system S,$O$,E 
 is still in the pure state even at $t\gg t_d$, and there is  IT observable
$B$ analogous to (\ref {AA5})
  which proves it.  Therefore in standard QM framework
it's incorrect to claim that IGUS percepts random events. 
The regarded IGUS model doesn't differs principally
from our MS scheme,
hence in Algebraic formalism IGUS  subjective perception  is described
by  $O$ restricted  state $\xi^O$ defined on $\cal {U}_O$
 which describes the random
outcomes for the input pure S state.  Consequently 
Algebraic QM  application 
  to Zurek IGUS model leads to the results which are  equivalent to
 Existential Interpretation of QM.

From the formal point of view
 all the   experiments in Physics  at the
final stage include  the human subjective perception
which simulated by $O$ state in our model.
The possible importance of observer in the quantum measurement
process was discussed first by London and Bauer (London,1939). They supposed
that  Observer Consciousness (OC) due to 'introspection
action' violates in fact Schr$\ddot{o}$dinger equation  
and results in the state reduction.
In our  theory  the process  of $O$  perception      
 doesn't violate  Schrodinger evolution 
relative to external RF $O'$.
In principle,
Self-description Theory  permits  to regard the relation between
MS, IGUS $O$ states and $O$ subjective information (impression), and
 one can try to extend it on the human perception.
 This is the separate, important problem which is beyond
our scope and here
we consider briefly only some its principal points.
Basing on  obtained results, our approach to it 
formulated here as the semiqualitative  Impression Model (IM). 
Following Algebraic QM  approach, IM assumes that
 $O$ perception affected only by 
$O$ internal states defined on $O$ observables subalgebra $\cal {U}_O$. 
For  $O$ perception   the  calibration assumption
introduced:  for any Q eigenstate $|s_i\rangle$ 
 after S measurement finished at $t>t_1$ 
and $O$ 'internal pointer'state is $|O_i\rangle$ observer $O$
 have the definite impression $J^O$ 
corresponding  to $J^O_i=q^O_i$ eigenvalue  - the information 
pattern percepted by $O$.
It  settles the hypothetical correspondence between MS quantum
dynamics model and the  human perception. 
Impression $J^O$ is $O$ subjective information which isn't
dynamical parameter and  its introduction can't have
any influence on  the theory dynamics.
 Furthermore, it's sensible to assume that
  if S state is the superposition $\psi_s$  
 then its measurement by $O$ also  results in appearance
for each individual event $n$ of some 
 definite  and unambiguous $O$ impression -
the information pattern  $J^O=\{q_l^{sup}(n)\}$
which is expressed as some finite sequence of real numbers. 
 In Algebraic formalism the corresponding
  $O$ subjective information - impression in the individual
event  is equal to  $J^O(n)=q^O_i$ and can be consistently defined
in this ansatz for our IM as the stochastic state 
appearing  with probability $|a_i|^2$.
In Algebraic QM  framework they corresponds to the restricted $O$ AIS $\xi^O_i$
defined on $\cal{U}_O$.
Eventually  in this approach IM doesn't need to exploit
  Von Neuman psychophysical parallelism hypothesis
(Von Neumann,1933).

To conclude, the quantum measurements were studied within the
 Information-Theoretical framework and  the self-description
 restrictions  on the information acquisition
are shown to be important in Quantum Measurement Theory.
Algebraic QM represents
the appropriate formalism of systems self-description,  
 in particular, 
 IGUS  $O$ observable algebra $\cal {U}_O$ defines $O$ restricted states
$\xi^O$ set $\Omega_O$.
The appearance of stochastic events 
 stipulated by  MS individual states restriction to  $\xi^O$ states
and results in the  state collapse observation by $O$. 
The regarded IGUS model is quite simple and on the whole doesn't
permit us to make any final conclusions at this stage.
Yet  the obtained results evidences
 that the  IGUS information
 restrictions and its interactions with the observed system 
should be accounted in Quantum  Measurement Problem analysis  (Zurek,1998).
\\
\\
\qquad \qquad \qquad {\Large { References}}\\
\\
 (1981) Y.Aharonov, D.Z. Albert Phys. Rev. D24, 359 
\\
 (2000) G.Bene, quant-ph 0008128
\\
(2003) Ph.Blanchard, P.Lugiewicz, R.Olkiewicz Phys. Lett A314, 29
\\
 (1979) O.Bratteli, D.Robinson 'Operators Algebra and
Quantum Statistical Mechanics' (Springer-Verlag, Berlin)
\\
 (1996) T.Breuer, Phyl. of Science 62, 197 (1995),
 Synthese 107, 1 (1996)
\\
 (1996) P.Busch, P.Lahti, P.Mittelstaedt,
'Quantum Theory of Measurements' (Springer-Verlag, Berlin,1996)
\\
 (1990) W. D'Espagnat, Found Phys. 20,1157,(1990)
\\
%
 (2002) R.Duvenhage, Found. Phys. 32, 1799  
\\
 (1994) A.Elby, J.Bub Phys. Rev. A49, 4213 
\\
 (1972) G.Emch, 'Algebraic Methods in Statistical Physics and
Quantum Mechanics',\\
 (Wiley,N-Y) 
\\
 (1988) D.Finkelstein, 'The Universal Turing Machine:
 A Half Century Survey', (ed. R.Herken, University Press, Oxford) 
\\
 (1996) D.Guilini et al., 'Decoherence and Appearance of
Classical World', (Springer-Verlag,Berlin) 
\\
 (1985) S.Kochen 'Symposium on Foundations of Modern Physics'
  , (World scientific, Singapour)
\\
(1968) J.M.Jauch 'Foundations of Quantum Mechanics' 
(Adison-Wesly, Reading)
\\
 (1990) P. Lahti Int. J. Theor. Phys. 29, 339 
\\
 (1939)  London F., Bauer E. 'La theorie de l'Observation'
 (Hermann, Paris)   
\\
 (1998) S.Mayburov, Int. Journ. Theor. Phys. 37, 401 
\\
 (2001) S.Mayburov  Proc. V  QMCC Conference, Capri, 2000,
(Kluwer, N-Y); $ \setminus$quant-ph 0103161
\\
 (2004) S.Mayburov Int. J.Theor. Phys,(to appear);
 Quant-ph 0205024; Quant-ph 0212099
\\
 (2005) S.Mayburov, Proc. of 'Foundations of Physics and Computations'
conference, Udine, 2004 (to appear )
\\
 (1998) P.Mittelstaedt 'Interpretation of
Quantum Mechanics and Quantum Measurement\\ Problem',
(Oxford Press, Oxford)
\\
 (1932) J. von Neuman 'Matematische Grunlanden
 der Quantenmechaniks' , (Berlin)
\\
 (1983) H.Primas,  'Quantum Mechanics,
 Chemistry and Reductionism' (Springer, Berlin)
 (1990) H.Primas,  in  'Sixty two years of uncertainty'
,ed. E.Muller, (Plenum, N-Y)
\\
 (1995) C. Rovelli, Int. Journ. Theor. Phys. 35, 1637; 
quant-ph 9609002  
\\
 (1947) I.Segal, Ann. Math., 48, 930    
\\
 (1993) K.Svozil 'Randomness and undecidability in Physics',
(World Scientific, Singapour)
\\
%
 (1961) E.Wigner,  'Scientist speculates',(Heinemann, London)
\\
 (1982) W.Zurek, Phys Rev, D26,1862 
\\
 (1998) W.Zurek Phys. Scripta , T76 , 186 
\\
\\
\end {document}